\documentclass[aps,prl,reprint,superscriptaddress,floatfix]{revtex4-2}
\usepackage{amsmath,amssymb}
\usepackage{graphicx}
\usepackage{bm}
\usepackage{hyperref}

\hypersetup{
    colorlinks=true,
    linkcolor=blue,
    citecolor=blue,
    urlcolor=blue,
    breaklinks=true
}

\begin{document}

\title{Stabilization of Interband Phase Solitons in Two-Band
Noncentrosymmetric Superconducting Rings}
\author{Yuriy Yerin}
\affiliation{Centro de F\'isica de Materiales (CFM-MPC),
Centro Mixto CSIC-UPV/EHU, E-20018 San Sebasti\'an, Spain}

\author{Boris Malomed}
\affiliation{Department of Physical Electronics, School of Electrical and Computer Engineering, Faculty of Engineering, Tel Aviv University, Tel Aviv 69978, Israel}
\affiliation{Instituto de Alta Investigaci\'on, Universidad de Tarapac\'a, Casilla 7D, Arica, Chile}

\author{Stefan-Ludwig Drechsler}
\affiliation{Institute for Theoretical Solid State Physics, IFW Dresden, 01069 Dresden, Germany}

\author{F. Sebastian Bergeret}
\affiliation{Centro de F\'isica de Materiales (CFM-MPC),
Centro Mixto CSIC-UPV/EHU, E-20018 San Sebasti\'an, Spain}
\affiliation{Donostia International Physics Center (DIPC),
E-20018 San Sebasti\'an, Spain}

\date{\today}

\begin{abstract}
Two-band superconductors maintain a relative interband phase which can carry winding soliton excitations in a superconducting ring, supported by independent winding numbers in the two bands. In rings of superconductors obeying the inversion symmetry the interband phase solitons are metastable states, separated from the uniform ground state by the energy of screening currents. In this work we find that, breaking the inversion symmetry strongly enough, one can make the soliton a true ground state. In that case, a magneto-electric coupling, absent in centrosymmetric materials, contributes critically above a certain threshold, a relevant, free-energy term, odd with respect to the winding number, which biases the energy balance in favor of a particular winding sign. Once the bias outweighs the energy cost that originally made the soliton metastable, a phase soliton with a finite winding number becomes the ground state, with chirality set by the applied field. In current--flux measurements performed in equilibrium states, the effect is demonstrated by field-odd soliton branches, that replace the metastable ones existing in mesoscopic rings, built by two-component superconductors, realizing a magneto-electric analog of the Little--Parks fluxoid-branch physics in the interband relative-phase sector.
\end{abstract}

\maketitle

\textit{Introduction.}--- Forcing the macroscopic quantum phase of a superconductor to wind around a ring is the basis of the flux quantization and the Little--Parks effect (periodic oscillations of the superconductivity transition temperature $T_{c}$ as a function of the magnetic flux threading the ring) \cite{LP_origin}. Superconductors with two order parameters---from MgB$_{2}$ to iron-based and and many other two-band materials---carry the second, internal phase degree of freedom, \textit{viz}., the relative phase between the two order parameters, which has no counterpart in conventional single-order-parameter superconductors. The internal phase supports a variety of topological textures beyond ordinary vortices, including
fractional-flux vortices, interband phase solitons (IBPSs), and multiple-$q$ states, in which the order parameter carries several coexisting winding wavevectors~\cite%
{Tanaka2002,Babaev2002,Gurevich2003, Samokhin_soliton, Kuplevakhsky2011, Babaev_3band, Tanaka2010, Lin2012, Vakaryuk2012, Malomed2015, Yerin_3band, Yerin2023}. Small oscillations of this phase form the Leggett mode, gapped by the interband coupling~\cite{Leggett1966}. By contrast, an IBPS in a superconducting ring is a nonlinear winding texture of the relative phase: if the two order parameters carry different winding numbers in the ring, the relative phase absorbs the mismatch between them. Such solitons feature a definite handedness (the sign of the winding number), making them potentially useful objects for the use in nonreciprocal and topological superconducting devices.
Thus, in a ring-shaped two-band superconductor with angular coordinate $\varphi $, which maintains two order parameters with phases $\chi _{1,2}$, carrying mutually independent integer winding numbers $n_{1,2}$, the relative phase, $\phi =\chi _{1}-\chi _{2}$, obeys the winding boundary condition
\begin{equation}
\phi (\varphi +2\pi )-\phi (\varphi )=2\pi \left( n_{1}-n_{2}\right) \equiv
2\pi n,  \label{eq:winding}
\end{equation}%
with integer $n$ being the interband winding number.

The potential use of IBPSs is impeded by the fact that they are metastable states in a mesoscopic ring of a centrosymmetric two-band superconductor. Indeed, the energy cost of the screening currents places the IBPS above the energy level of the spatially uniform state. Therefore, IBPSs can be created dynamically, but not realized as the global ground state~\cite{Bluhm2006,Kuplevakhsky2011}.  Whether an interband soliton could ever be a ground state, rather than a merely metastable configuration, has remained open since its original prediction.

In this work, we show that breaking inversion symmetry, as in noncentrosymmetric superconductors, provides a route to making the IBPS the ground state. An applied field activates a magneto-electric coupling—described by a generalized Lifshitz invariant—that biases the two winding directions unequally: it adds energies of opposite sign to the IBPS states of opposite handedness (i.e., opposite winding-number sign). Once the negative energy shift outweighs the screening energy cost, the soliton becomes the ground state, with the handedness flipping when the field is reversed. Because the underlying winding number can be detected in scanning-SQUID measurements of mesoscopic rings build by two-component noncentrosymmetric superconductors ~\cite{Bluhm2006}, the prediction is directly testable: the equilibrium current--flux response should show the field-odd soliton branch, in place of the previously observed metastable one. 

The analogy to the conventional Little--Parks effect occurs at the level of fluxoid branches: in the conventional effect, the magnetic flux selects the winding number of the common superconducting phase and modulates $T_{c}$ periodically, whereas here the Lifshitz invariant biases the interband (relative-phase) winding number Eq. (\ref{eq:winding}), deep in the superconducting state.

\begin{figure}[t]
\centering
\includegraphics[width=\columnwidth]{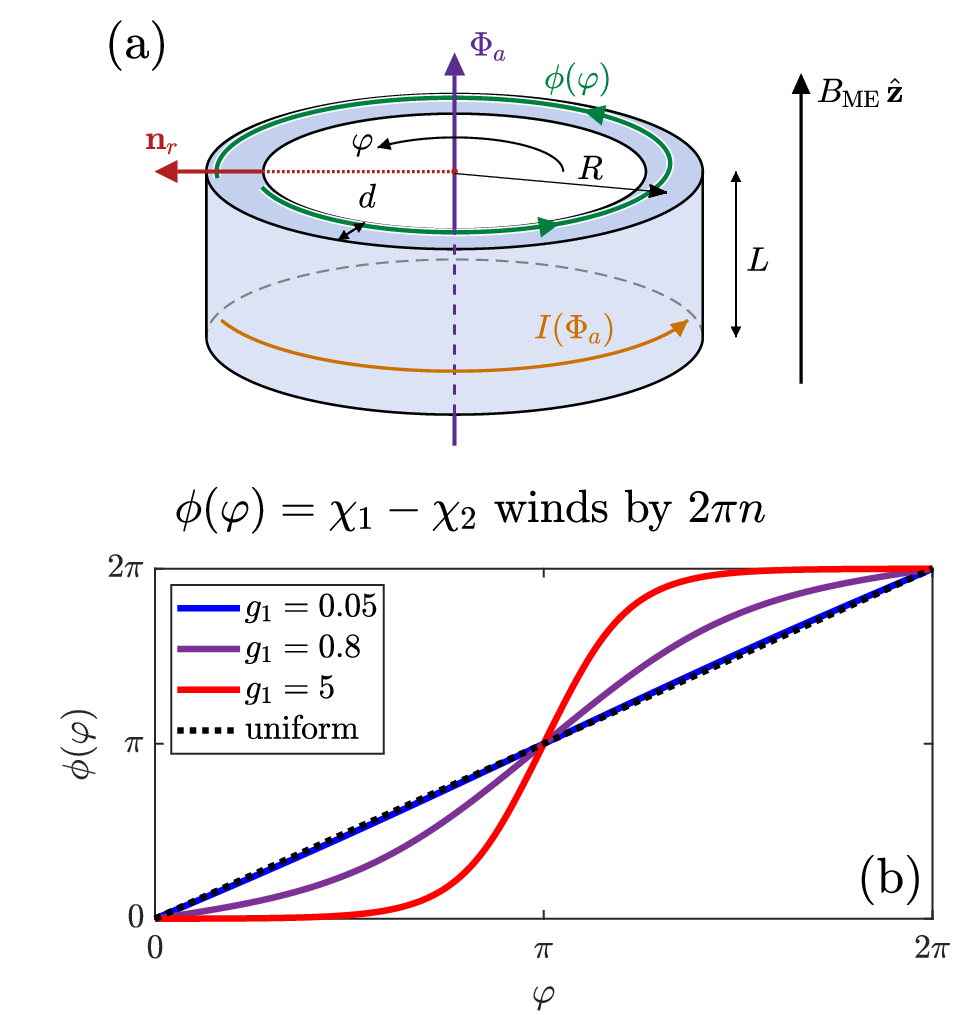}
\caption{(a) The physical setup: a thin-walled ring (with the mean radius $R$, wall thickness $d$, and height $L$) of a noncentrosymmetric two-band superconductor, whose local inversion-symmetry-breaking axis $\mathbf{n}_{r}$ is radial (red arrows; the dotted line: a radius). The green curve shows the relative-phase profile: its radial displacement represents $\protect\phi (\protect\varphi )$ (with the phase circulation $2\protect\pi $ around the ring) for visualization only, not a real-space deformation [cf.\
(b)]. The axial bias magnetic field $B_{\mathrm{ME}}$ activates the Lifshitz invariant. The probe flux $\Phi _{a}$ is tuned independently and the circulating current $I(\Phi _{a})$ is to be measured. (b) An exact double-sine-Gordon profiles $\protect\phi (\protect\varphi )$ with $n=1$ [see Eq.~\eqref{eq:winding}] at increasing values of the locking parameter $g_{1}$ ($g_{2}=g_{1}/8$, in units of $K_{\protect\phi }$), from the nearly linear profiles to the narrow kink.}
\label{fig:setup}
\end{figure}

\textit{The Ginzburg-Landau functional and London reduction.}--- As shown in Fig. \ref{fig:setup}, we consider a thin-walled two-band ring, modelled as a cylinder with mean radius $R$, wall thickness $d$, and height $L$ along the cylinder axis $\hat{\mathbf{z}}$ (parallel to the uniform magneto-electric bias $B_{\mathrm{ME}}$), with $d\lesssim \xi _{1,2}\ll R$, where $\xi _{1,2}$ are coherence lengths for each band, neglecting amplitude variations across the wall (i.e., along the radial direction) to leading order with the value at the mean radius $A_{\varphi}(\varphi )\equiv A_{\varphi }(R,\varphi )$. The local scaled azimuthal gauge field, which as a term in the covariant derivative $D_{\varphi }$, is
\begin{equation}
\mathcal{A}(\varphi )\equiv \frac{2\pi RA_{\varphi }(\varphi )}{\Phi _{0}},\qquad D_{\varphi }=\partial _{\varphi }-i\mathcal{A}(\varphi ),
\label{eq:calA_def}
\end{equation}%
where $\Phi _{0}=hc/(2|e|)$ 
We use the cylindrical gauge, with $\bm %
A=A_{\varphi }\,\hat{\bm\varphi }$ and $\partial _{\varphi }A_{\varphi }=0$ at the mean-radius position. Other field components do not couple to the azimuthal phase.

We start from the standard two-band Ginzburg-Landau (GL) functional per
unit length~\cite{Tanaka2002,Gurevich2003,Kuplevakhsky2011}, augmented by the band-resolved magneto-electric Lifshitz term allowed by the broken inversion symmetry~\cite{MineevSamokhin2008,Samokhin2013},
\begin{align}
& \mathcal{G}=\int_{0}^{2\pi }\!\mathrm{d}\varphi \Bigg\{\sum_{a=1,2}\left[
a_{a}|\Psi _{a}|^{2}+\frac{b_{a}}{2}|\Psi _{a}|^{4}+\frac{K_{a}}{R^{2}}%
|D_{\varphi }\Psi _{a}|^{2}\right]   \notag \\
& \quad -\Gamma _{1}(\Psi _{1}^{\ast }\Psi _{2}+\mathrm{c.c.})-\Gamma _{2}%
\big[(\Psi _{1}^{\ast }\Psi _{2})^{2}+\mathrm{c.c.}\big]  \notag \\
& \quad +\sum_{a=1,2}\kappa _{a}B_{z}\,\mathrm{Im}\big(\Psi _{a}^{\ast
}D_{\varphi }\Psi _{a}\big)\Bigg\}+\frac{1}{8\pi }\int \mathrm{d}^{2}r\,\big(%
\bm h-\bm H\big)^{2},  \label{eq:GL_full}
\end{align}%
where $\Psi_{a}$ ($a=1,2$) are the complex order parameters of the two bands, $a_{a}$, $b_{a}$, $K_{a}$ are the intraband coefficients with $a=1,2$, $\Gamma _{1}$ the Josephson interband coupling, and $\Gamma _{2}$ the coefficient of the second-harmonic interband locking term~\cite{Stanev, Yerin_magneto, Yerin2023}. The term with the Lifshitz coupling $\kappa _{a}$ is the band-resolved magneto-electric Lifshitz invariant, with the uniform axial magnetic field $%
B_{z}\equiv B_{\mathrm{ME}}$ representing the magneto-electric coupling, which is evaluated at the superconducting annulus. The gauge-invariant current bilinear is $\mathbf{j}_{a}=\mathrm{Im}(\Psi _{a}^{\ast}D_{\varphi }\Psi _{a})$, whose product
with the axial field is even and odd with respect to the time reversal and spatial inversion, respectively, It is therefore forbidden in centrosymmetric crystals and becomes symmetry allowed when the inversion symmetry is broken \cite{MineevSamokhin2008,Samokhin2013}. For the radial-Rashba geometry~\cite{Rashba1960,BychkovRashba1984} of Fig.~\ref{fig:setup}(a) the symmetry-allowed invariant is $\kappa _{a}(\hat{\mathbf{r}}\times \mathbf{B})\cdot
\mathbf{j}_{a}$; with $\mathbf{B}=B_{z}\hat{\mathbf{z}}$ and $\hat{\mathbf{r}}$ the radial unit vector, one has $(\hat{%
\mathbf{r}}\times \hat{\mathbf{z}})\cdot \hat{\bm\varphi }\equiv -1$, therefore the invariant reduces, up to a sign absorbed into $\kappa _{a}$, to the respective term in Eq.~\eqref{eq:GL_full}.

For a thin-walled ring the order parameters are essentially rigid, so that in the present geometry they depend only on the azimuthal coordinate, $\Psi_{a}(\varphi )=\Delta _{a}(\varphi )e^{i\chi _{a}(\varphi )}$, with $\Delta_{a}(\varphi )$ considered as constants, the variables being phases $\chi_{a}(\varphi )$. The relative phase $\phi =\chi _{1}-\chi _{2}$ and the stiffness-weighted common phase $\chi =c_{1}\chi _{1}+c_{2}\chi _{2}$ (with weights $c_{a}=\rho _{a}/\left( \rho _{1}+\rho _{2}\right) $ and phase stiffnesses $\rho _{a}=2K_{a}\Delta _{a}^{2}/R^{2}$) diagonalize the local
gradient energy, and the Lifshitz invariant splits into common and relative parts, $\eta_{c}=\eta _{1}+\eta _{2}$ and $\eta _{\phi }=(\eta _{1}\rho _{2}-\eta_{2}\rho _{1})/\rho $, respectively, with $\eta _{a}\equiv \kappa_{a}B_{z}\Delta _{a}^{2}$. The common part produces only a rigid flux offset, $f_{L}=\eta _{c}/\rho $, whereas the relative part biases the interband soliton. Crucially, $\eta _{\phi }$ vanishes unless the band-normalized responses are unequal, $\eta _{1}/\rho _{1}\neq \eta_{2}/\rho _{2}$, hence the bias is a genuine multiband effect, that broken inversion symmetry alone does not produce.

Reduction of the GL functional, detailed in the Supplemental
Material~\cite{SM}, leads to the Gibbs energy,
\begin{equation}
g(n;f_{H})=(1-\epsilon )\,h^{2}(f_{H}-nc_{2}+f_{L})+Y_{\mathrm{%
sol}}^{\mathrm{DSG}}(|n|)+\Lambda n,  \label{eq:g_total}
\end{equation}%
which is a central point of this work. It is a function of the winding number $n$ and reduced orbital flux, $f_{H}\equiv \left[ \Phi _{\mathrm{ME}}(B_{\mathrm{ME}})+\Phi _{a}\right] /\Phi _{0}$, with the numerator being the total flux through the ring, whose constant bias part is absorbable into the periodic one, so that the variation of $\Phi _{a}$ affects only $f_{H}$, and $h(x)\equiv x-\lfloor x\rceil $ is the signed deviation of $x$ from the nearest integer. The first term in $g$ is the common-mode electromagnetic Gibbs energy (with the screening parameter of the thin-walled ring $\epsilon \equiv dR/2\lambda ^{2}\ll 1$, and the flux shifted by $f_{L}$). The second term $Y_{\mathrm{sol}}^{\mathrm{DSG}}(|n|)$ Eq. (\ref{eq:g_total}) is the self-energy of the double sine-Gordon (DSG) soliton, \textit{viz}., the cost of forcing $\phi $ to advance by $2\pi |n|$ around the ring against the chirality-even interband locking potential
\begin{equation}
V(\phi )=-g_{1}\cos \phi -g_{2}\cos (2\phi ),  
\label{V}
\end{equation}
[$g_{1}\equiv 2\Gamma _{1}\Delta _{1}\Delta _{2}$, $g_{2}\equiv 2\Gamma
_{2}\Delta _{1}^{2}\Delta _{2}^{2}$], measured from the homogeneous minimum, so that $Y_{\mathrm{sol}}^{\mathrm{DSG}}(0)=0$. For weak locking the cost is purely elastic: in the limit $g_{1,2}\to 0$, $Y_{\mathrm{sol}}^{\mathrm{DSG}}(|n|)\rightarrow \kappa_{\phi }|n|^{2}$, where $\kappa _{\phi }=\pi K_{\phi }/E_{\ast }$ and $K_{\phi }=\rho _{1}\rho_{2}/\rho $; for the centrosymmetric normalization of Ref.~\cite{Kuplevakhsky2011}, used in the figures here, $\kappa _{\phi }=c_{1}c_{2}$.
For strong locking the winding concentrates into a localized kink, whose energy is governed by the full depth of the locking potential rather than by the elastic stiffness alone. The third term in Eq. (\ref{eq:g_total}) is the topological Lifshitz bias
\begin{equation}
\Lambda \equiv \frac{2\pi \eta _{\phi }}{E_{\ast }},\qquad E_{\ast }=\frac{\Phi _{0}^{2}\epsilon }{2L_{m}},\qquad L_{m}=\frac{4\pi ^{2}R^{2}}{L},
\label{eq:Lambda}
\end{equation}
which is odd with respect to the winding number, thereby selecting the chirality (handedness). Here $E_{\ast }$ is the electromagnetic energy scale and $L_{m} $ the geometric inductance of the ring of height $L$. The soliton is a stationary solution of the DSG equation obtained by variation of the relative-phase functional, for which the relative phase advances by $2\pi |n|$ around the ring. An elliptic-function form of this solution, its ring quantization, the quadrature construction and branch selection used for $g_{2}<0$, and the associated Leggett mode are given in Supplemental Material~\cite{SM}.

\textit{The stability criterion and flux branches.}--- The relative Lifshitz contribution alone splits the two chiralities by $2\Lambda n$. When the conventional flux bias is compensated, so that $f_{H}+f_{L}=0$ modulo an integer, the electromagnetic terms are symmetric with respect to $n\rightarrow -n$, the full branch splitting, being $g(n,f_{H})-g(-n,f_{H})=2\Lambda n$. For $\Lambda >0$ the bias favors negative winding numbers, and the rival of
the spatially uniform state, with $n=0$, in the competition for the role of the ground state, is the state with negative integer $n=-k$ ($k\geq 1$). The competing terms become degenerate at a threshold value of the bias, $\Lambda_{\mathrm{GS}}(k;f_{H})$, fixed by equating their respective values of the reduced Gibbs energy, $g(-k;f_{H})=g(0;f_{H})$. Therefore, the finite-winding ground state exists if $\Lambda $ exceeds the threshold value,
\begin{equation}
\begin{split}
\Lambda & >\min_{k\geq 1}\Lambda _{\mathrm{GS}}(k;f_{H}),\qquad \Lambda _{%
\mathrm{GS}}=\frac{1}{k}\Big\{Y_{\mathrm{sol}}^{\mathrm{DSG}}(k) \\
& +(1-\epsilon )\big[h^{2}(f_{H}+kc_{2}+f_{L})-h^{2}(f_{H}+f_{L})\big]\Big\},
\end{split}
\label{eq:GS_condition}
\end{equation}%
which is obtained from Eq.~\eqref{eq:g_total} with $Y_{\mathrm{sol}}^{%
\mathrm{DSG}}(0)=0$. The electromagnetic term raises or lowers the threshold through the flux compensation, while the topological stabilization of the IBPS is provided by $\Lambda n$. With the growth of $\Lambda $, increasingly negative winding numbers may become energetically favorable, with each transition determined by the corresponding difference of the soliton and electromagnetic energies. This result is somewhat similar to the recently predicted possibility of replacing the ordinary ground state by one in the form of a vortex mode in the two-dimensional binary Bose-Einstein condensate
under the combined action of the spin-orbit coupling and gradient magnetic field~\cite{Luo2024}.

\begin{figure}[t]
\centering
\includegraphics[width=\columnwidth]{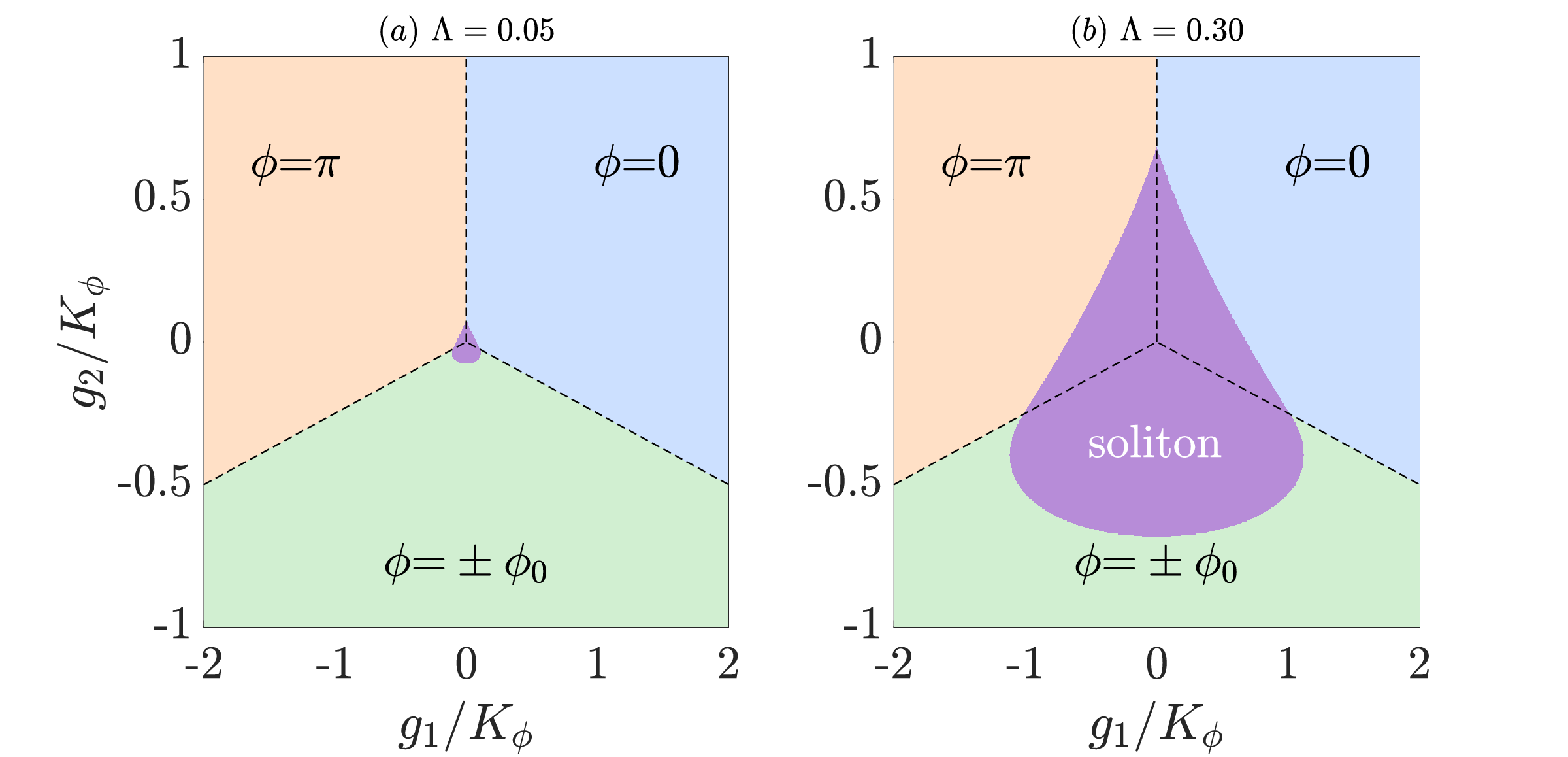}
\caption{The ground-state phase diagram in the plane of interband-coupling parameters, $(g_{1}/K_{\protect\phi },g_{2}/K_{\protect\phi })$, as produced by the minimization of energy Eq.~\eqref{eq:g_total} with respect to integer $n$ at $f_{H}=1/2$. Four domains appear: the homogeneous states $\protect\phi =0$ (blue) and $\protect\phi =\protect\pi $ (orange), the double-well state $\protect\phi =\pm \protect\phi _{0}$ (green), and the winding interband soliton (purple). Dashed lines: homogeneous-state boundaries ($g_{1}=0$ for $g_{2}>0$; $g_{2}=-|g_{1}|/4$ for $g_{2}<0$); spinodal branches ($|g_{1}|=4|g_{2}|$) are not shown. As $\Lambda $ grows,
the area of the soliton ground state expands from a small region near the origin (a)~$\Lambda =0.05$ to a broad basin (b)~$\Lambda =0.30$. Parameters: $c_{2}=0.5$, $\protect\epsilon =0.05$, $f_{L}=0$; locking strengths in units of $K_{\protect\phi }$, with $\protect\kappa _{\protect\phi }=c_{1}c_{2}$. }
\label{fig:phase_diagram}
\end{figure}

\textit{Homogeneous phases and the soliton phase diagram.}--- The analysis of the energy functional given by Eq.~\eqref{eq:g_total} leads to a ground-state phase diagram in the plane of interband-coupling parameters $\left( g_{1},g_{2}\right) $ [see Eq. (\ref{V})], plotted in Fig.~\ref{fig:phase_diagram} . The spatially uniform global minima of $V(\phi )$ are $\phi =0$, $\phi =\pi $, or---for $g_{2}<0$ and $|g_{1}|<-4g_{2}$---a degenerate double-well pair, $\phi =\pm \phi _{0}$ with $\cos \phi_{0}=-g_{1}/4g_{2}$. The boundaries between them are $g_{1}=0$ (separating $\phi =0$ and $\phi =\pi $ for $g_{2}>0$) and lines $g_{2}=-|g_{1}|/4$ (bounding the $\pm \phi _{0}$ domain for $g_{2}<0$). The upper branches $|g_{1}|=4|g_{2}|$ at $g_{2}>0$ are spinodal lines of local stability, rather than ground-state boundaries. At fixed values of $\Lambda ,f_{L},c_{2},f_{H}$, we minimize the energy~\eqref{eq:g_total} with respect to the integer $n$: if the minimum is at $n=0$, the state is the spatially uniform phase minimizing $V$, and the minimum at $n\neq 0$ corresponds to a finite-winding soliton. This analysis produces four domains: the uniform states with phases $0$, $\pi $, $\pm \phi_{0}$, and, separately, the soliton region. The character of the soliton within this region is set by the kink's angular width (in units of $R$)  $\ell _{\varphi }=\sqrt{K_{\phi }/\Delta V}$---fixed by the ratio of the relative-phase stiffness $K_{\phi }$ to the locking-potential depth $\Delta V=V_{\max }-V_{\min }$---compared with the full circuit $2\pi $. For weak locking, $\ell _{\varphi }\gg 2\pi $ and the state is a delocalized winding of the relative-phase momentum, whose energy cost is given by the elastic term $\kappa _{\phi }|n|^{2}$, whereas for strong locking $\ell _{\varphi }\ll 2\pi $ and the winding pattern contracts to a localized kink. The switch between these regimes is a smooth crossover, rather than a sharp phase transition~\cite{SM}. It explains the shape of the phase diagram in Fig. \ref{fig:phase_diagram}: the elastic-energy cost is far cheaper than the potential-barrier cost, defined by Eq. (\ref{V}), therefore the soliton basin in the diagram is centered where the locking vanishes, and widens as $\Lambda $ grows.

\textit{Discussion.}--- The variation of the relative-phase-controling
parameter $\Lambda n$ leaves the DSG soliton equation unchanged for uniform $\eta _{\phi }$ yet reorders the winding states globally. This is why a finite-winding state can become the ground state even though the centrosymmetric model supports only metastable soliton branches \cite{Kuplevakhsky2011}.

\begin{figure}[t]
\centering
\includegraphics[width=\columnwidth]{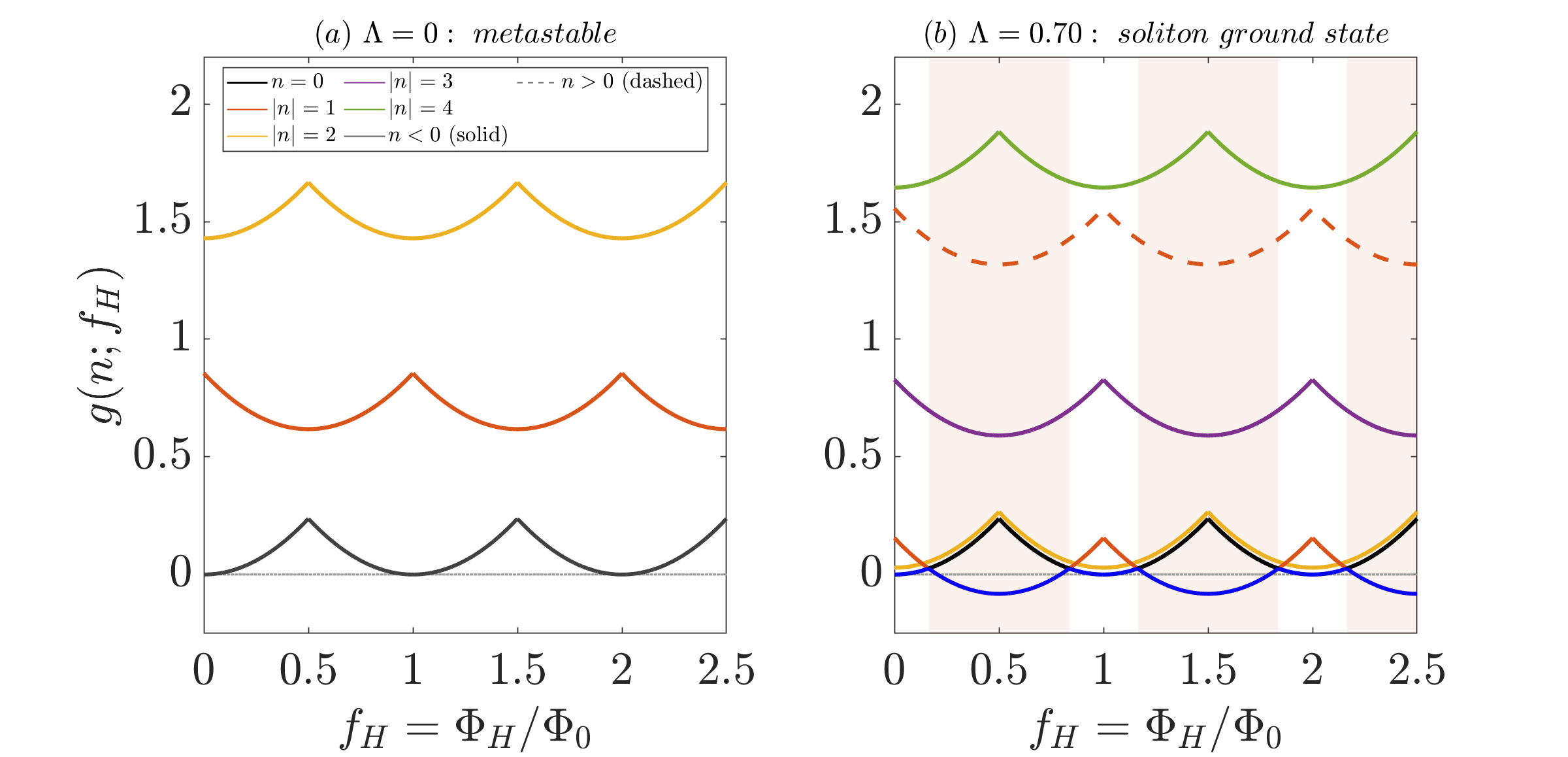}
\caption{The reduced Gibbs energy $g(n;f_{H})$, defined as per Eq.~
\eqref{eq:g_total}, vs. the applied flux for several winding sectors, with the line style encoding the sign of $n$ (solid for $n<0$, dashed for $n>0$). The bold black and blue curves represent the ground state for (a) $\Lambda =0$ and (b) $\Lambda =0.7$, respectively. Shaded windows mark flux ranges where the solitons pick up the role of the ground state. The parameters are $g_{1}=0.8$, $g_{2}=0.1$; the others are as in Fig.~\protect\ref{fig:phase_diagram}.}
\label{fig:gibbs_flux}
\end{figure}

Figure~\ref{fig:gibbs_flux} makes this explicit. At $\Lambda =0$ [Fig.~\ref{fig:gibbs_flux}(a)] every finite-$|n|$ curve lies above the black curve for the uniform state with $n=0$, recovering the centrosymmetric metastability of Ref.~\cite{Kuplevakhsky2011}. A finite bias adds the chirality splitting $2\Lambda |n|$. For $\Lambda =0.7$ [Fig.~\ref{fig:gibbs_flux}(b)] the curve for $n=-1$ (orange solid) \emph{drops below} the $n=0$ curve in the shaded flux windows, picking up the role of the ground state, while its $n=1$
partner (orange dashed) is pushed up towards instability. Thus, the ground state switches between the uniform ($n=0$) and solitonic ($n=-1$) states as the flux is varied. The blue curve in Fig.~\ref{fig:gibbs_flux}(b) further highlights the switching.

The phase diagram in Fig.~\ref{fig:phase_diagram} provides a complementary view at the fixed flux $f_{H}=1/2$. The soliton region is centered at the origin $g_{1}=g_{2}=0$, where the interband locking vanishes: the relative phase is free there, a finite-winding state costs only its kinetic energy $c_{1}c_{2}n^{2}$, and the Little--Parks term renders the $n=0$ and $n=-1$ fluxoid sectors nearly degenerate. The nonzero Lifshitz term lifts the degeneracy, so that even small $\Lambda $ stabilizes the soliton state. Away from the origin, the stronger locking raises the winding cost, requiring larger $\Lambda $ to offset it, and, accordingly, the soliton basin expands outward
as $\Lambda $ grows [Figs.~\ref{fig:phase_diagram}(a,b)]. The mechanism is not tied to the symmetric choice $c_{2}=1/2$ fixed in the figures: for $c_{2}\neq 1/2$ the same finite-winding region appears, shifted with respect to the flux, as it follows from Eq.~\eqref{eq:g_total}.

\begin{figure}[t]
\centering
\includegraphics[width=\columnwidth]{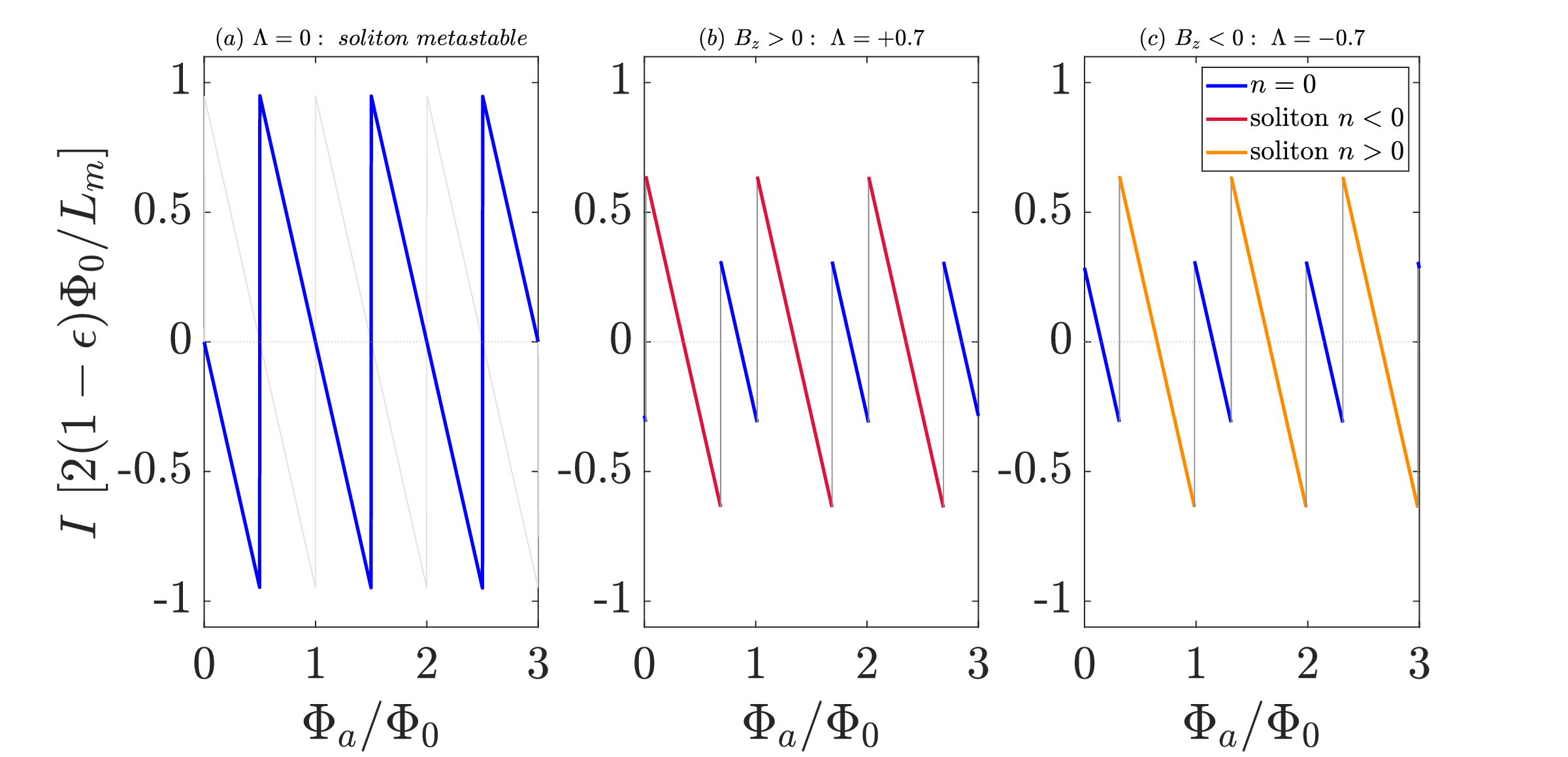}
\caption{The equilibrium current $I_{n}(f_{H})=-\partial \mathcal{G}%
/\partial \Phi _{a}$ versus the probe flux at a fixed magneto-electric bias. Occupied sectors are drawn bold [blue for $n=0$, and crimson or orange for the IBPS with $n<0$ or $n>0$ , respectively. Faint branches in (a) are metastable ones. Thin vertical lines mark fluxoid jumps, and the dotted line stands for $I=0$. (a) $\Lambda =0$: the equilibrium response is similar to the $n=0$ sawtooth in Ref.~\protect\cite{Bluhm2006}. (b) $B_{\mathrm{ME}}>0$ and (c) $B_{\mathrm{ME}}<0$ ($\Lambda =\pm 0.7$): soliton segments appear in the equilibrium response. Their chirality and flux positions reverse with the reversal of the bias. Parameters are the same as in Fig.
\protect\ref{fig:gibbs_flux}, with $f_{L}=0.15\,\mathrm{sgn}(B_{\mathrm{ME}}) $ in (b,c).}
\label{fig:current_flux}
\end{figure}

\textit{A possible scheme for the experimental realization.}--- Phase-soliton physics with $n_{1}\neq n_{2}$ has already been observed in Josephson-coupled aluminum bilayer rings simulating a two-band superconductor, via scanning-SQUID measurement of the circulating current vs. the applied flux~\cite{Bluhm2006}. In that case, such states arise only dynamically---first, flux sweeps drive the weaker condensate through its fluxoid instability---and they survive only as kinetically trapped metastable states, absent in the spectrum of equilibrium modes. Our mechanism predicts the qualitatively different outcome in Fig.~\ref{fig:current_flux}: the Lifshitz term makes the IBPS the ground state, allowing it to appear in the thermally-averaged $I(\Phi _{a})$ curve, without the use of the dynamical sweep or trapping. We take $B_{\mathrm{ME}}$ for the spatially uniform bias in the annulus, which fixes the values of $(\eta_{a},f_{L},\Lambda )$, and vary flux $\Phi_{a}$ as the probe. The separation of physical scales is natural: one flux quantum corresponds to millitesla fields in a micron-size ring while $\Lambda \sim 1$ requires a Tesla-scale bias~\cite{SM}, so the probe sweep perturbs the Lifshitz parameters no more than at the percent level. The circulating current is then
\begin{equation}
I_{n}=-\frac{\partial \mathcal{G}}{\partial \Phi _{a}}\bigg|_{B_{\mathrm{ME}%
}}\propto -(1-\epsilon )\,h(f_{H}-nc_{2}+f_{L}),  \label{eq:current}
\end{equation}%
which at $\Lambda =0$ reduces to the single-order-parameter sawtooth discovered in Ref.~\cite{Bluhm2006}, but for $\Lambda \neq 0$ it contains soliton segments at irregularly-spaced, sub-$\Phi _{0}$-shifted transitions. The soliton branches that the authors of Ref.~\cite{Bluhm2006} could observe only as metastable states thus appear here as the ground state. Both Lifshitz terms are odd with respect to $B_{\mathrm{ME}}$, therefore reversing the bias implies $(\Lambda ,f_{L})\rightarrow (-\Lambda ,-f_{L})$. While $f_{L}$ alone merely shifts the flux origin, its joint reversal with $\Lambda $ moves the soliton segments to different flux positions and reverses their chirality [see Figs.~\ref{fig:current_flux}(b,c)]. In fact the reversal is
exact: the three bias-generated quantities $\Phi _{\mathrm{ME}}$, $f_{L}$, and $\Lambda $ are all odd with respect to $B_{\mathrm{ME}}\rightarrow -B_{\mathrm{ME}}$, hence, after recentering the probe-flux axis by the constant $\Phi _{\mathrm{ME}}/\Phi _{0}$ the reduced Gibbs energy obeys $g(n;f_{H})|_{-B_{\mathrm{ME}}}=g(-n;-f_{H})|_{+B_{\mathrm{ME}}}$ and the equilibrium current satisfies the point-inversion antisymmetry condition, $I(f_{H};-B_{\mathrm{ME}})=-I(-f_{H};+B_{\mathrm{ME}})$, which connects Figs.~\ref{fig:current_flux}(b) and (c). This argument suggests a sharp experimental test for spotting the expected IBPS ground state: bias-even contributions (depairing, heating, geometric flux offsets) cannot mimic it, nor can it be done by the chirality-symmetric metastable solitons reported
in Ref.~\cite{Bluhm2006}.

\begin{table}[t]
\caption{Illustrative magneto-electric bias-field scales for a soliton
ground state ($q_L\simeq0.6$, ring radius $R=1\,\protect\mu\mathrm{m}$), from the estimates of the Supplemental Material~\protect\cite{SM}; the underlying couplings are assumed ranges for each platform class, not material-specific fits.}
\label{tab:platforms}%
\begin{ruledtabular}
\begin{tabular}{ll}
Platform class & $B_{\rm ME}$ (T) \\
\hline
Heavy-fermion (e.g., CePt$_3$Si) & $0.1$--$3$ \\
Strong spin--orbit bulk (e.g., Li$_2$Pt$_3$B) & $0.7$--$7$ \\
Moderate spin--orbit bulk (e.g., Mo$_3$Al$_2$C) & $7$--$34$ \\
Oxide interface (e.g., LaAlO$_3$/SrTiO$_3$) & $0.1$--$3$ \\
\end{tabular}
\end{ruledtabular}
\end{table}

Introducing the scaled relative-phase winding, $q_{L}=\eta _{\phi }/K_{\phi} $, Eq.~\eqref{eq:Lambda} yields $\Lambda =2\kappa _{\phi }q_{L}$ ($\kappa_{\phi }=\pi K_{\phi }/E_{\ast }$). The IBPS ground state (with $\Lambda \sim 0.3$) then requires only $q_{L}\sim 0.6$, which is a fraction of the relative-phase twist around the ring, i.e., the helical pitch $2\pi R/q_{L}$ on the order of the ring's length~\cite{SM}.

The antisymmetric spin-orbit coupling that underlies the magneto-electric response is well established experimentally in noncentrosymmetric superconductors through violation of the Pauli limit, anomalous Knight shifts, and directly resolved Rashba band splittings~\cite{Smidman2017}, therefore the microscopic ingredients needed for the creation of a sizable Lifshitz invariant that should affect the relative phase, are available. Whether the band-selective combination reaches the level of $q_{L}\sim 0.6$ is material-specific; for micron-scale rings ($R\simeq 0.5$--$1\mathrm{~\mu }$m, wall thickness $d\lesssim \xi \sim 10$--$20$~nm) the order-of-magnitude
estimates given in the Supplemental Material~\cite{SM} yield the
platform-resolved bias ranges indicated in Table~\ref{tab:platforms}. The necessary probe may be the scanning-SQUID protocol outlined in Ref.~\cite{Bluhm2006}, with millitesla coil fields supplying the flux quanta. Because a fixed crystalline polar axis would make the Lifshitz coefficient oscillating around the ring, the bulk compounds shown in Table I serve primarily as benchmarks for attainable spin--orbit and electronic scales: realizing the azimuthally uniform radial-axis coupling in Fig.\ \ref{fig:setup} requires incorporating these ingredients into the engineered shell-shaped, curved-interface, or textured annular structures, rather than just planar single crystals~\cite{SM}. The Tesla-scale bias implies that these estimates can be compatible with the superconductivity in the ring geometry, particularly in the lower and intermediate ranges shown in Table~%
\ref{tab:platforms}: because the axial field is parallel to the thin wall, the orbital depairing across the wall is suppressed for $d\lesssim \xi $ [the parallel critical field scales as $\Phi _{0}/(\xi d)$], while the antisymmetric spin-orbit coupling can strongly modify the Pauli limit, as the measured examples of the Pauli-limit violation suggest. The same spin-orbit physics that enables the magneto-electric response can thus enlarge the field window in which the bias may be applied, although the quantitative limit is material- and geometry- dependent~\cite{SM}.

\textit{In summary}, a band-resolved magneto-electric Lifshitz invariant provides an equilibrium route to chirality-selective interband soliton physics in multiband superconducting doubly-connected geometries, converting the metastable solitons of centrosymmetric superconductors into a field-selected thermodynamic ground state.
Experimentally, the transition to the soliton ground state is revealed by finite-winding segments appearing directly in the equilibrium current--flux response, with their chirality and flux positions reversing with the magneto-electric bias. For favorable material platforms, the required bias fields are estimated to range from sub-Tesla to a few Tesla.

\textit{Acknowledgments.}
We thank Andrei Mazanik  for useful discussions. F.S.B  acknowledges financial support from the Spanish MCIN/AEI/10.13039/501100011033 through grant 
PID2023-148225NB-C31, and from the European Union’s Horizon Europe 
through grant JOSEPHINE (No. 101130224).

\clearpage
\setcounter{equation}{0}
\setcounter{figure}{0}
\setcounter{table}{0}
\setcounter{section}{0}
\renewcommand{\theequation}{S\arabic{equation}}
\renewcommand{\thefigure}{S\arabic{figure}}
\renewcommand{\thetable}{S\arabic{table}}

\begin{center}
{\textbf{\large Supplemental Material for\\
"Stabilization of Interband Phase Solitons in Two-Band Noncentrosymmetric Superconducting Rings"}}
\end{center}

This Supplemental Material provides: (i) the reduction of the two-band Ginzburg--Landau functional to the reduced Gibbs energy, including the kinetic diagonalization and the common/relative Lifshitz split; (ii) the closed elliptic form of the double--sine-Gordon running-soliton energy in the real-root regime, together with a quadrature-based construction in the complex-root regime, together with the branch-selection procedure used when
several winding branches coexist; (iii) the explicit elliptic expressions for the harmonic integrals $I_{1}$ and $I_{2}$ quoted in the main text; (iv) the crossover between a localized kink and a delocalized (finite relative-phase momentum) winding within the soliton region; (v) the small-amplitude (Leggett-mode) limit of the same relative-phase sector; (vi) the reduction to the single-harmonic sine-Gordon limit; and (vii) order-of-magnitude estimates of the magneto-electric field required for representative material platforms.

\section*{I. The London reduction and the reduced Gibbs energy}

Here we give the detailed reduction of the two-band Ginzburg--Landau functional from the main text to the reduced Gibbs energy.

Our analysis is performed in the framework of the London limit. Deep in the superconducting state the amplitudes are stiff, $\Delta _{a}(\varphi )\simeq \Delta _{a}$, and only the phases $\chi _{a}(\varphi )$ vary. Writing $\Psi _{a}=\Delta _{a}e^{i\chi _{a}}$, the gradient 
and the 
Lifshitz terms of the
energy functional per unit length in the main text become,
\begin{align}
\mathcal{G}=& \int_{0}^{2\pi }\!\mathrm{d}\varphi \Bigg[\sum_{a=1,2}\frac{%
\rho _{a}}{2}(\chi _{a}^{\prime }-\mathcal{A})^{2}+\sum_{a=1,2}\eta
_{a}(\chi _{a}^{\prime }-\mathcal{A})  \notag \\
& -g_{1}\cos \phi -g_{2}\cos 2\phi \Bigg]+\frac{1}{8\pi }\!\int \!\mathrm{d}%
^{2}r\,(\bm h-\bm H)^{2},  \label{eq:GL_London_SM}
\end{align}
with the phase stiffnesses $\rho _{a}=2K_{a}\Delta _{a}^{2}/R^{2}$,
the Lifshitz couplings $\eta _{a}=\kappa _{a}B_{z}\Delta _{a}^{2}$,
the interband locking strengths $g_{1}=2\Gamma _{1}\Delta _{1}\Delta _{2}$, $g_{2}=2\Gamma
_{2}\Delta _{1}^{2}\Delta _{2}^{2}$, and $\mathcal{A}$ as
the reduced vector potential. The relative phase is $\phi =\chi _{1}-\chi _{2}$.

To diagonalize the kinetic energy, we introduce the stiffness-weighted common phase $\chi =c_{1}\chi _{1}+c_{2}\chi _{2}$, with 
the
weights $c_{a}=\rho_{a}/\rho $ and $\rho =\rho _{1}+\rho _{2}$. Due to $c_{1}+c_{2}=1$, one has
\begin{equation}
\chi _{1}^{\prime }=\chi ^{\prime }+c_{2}\phi ^{\prime },\qquad \chi
_{2}^{\prime }=\chi ^{\prime }-c_{1}\phi ^{\prime }.
\end{equation}%
The kinetic term then separates exactly,
\begin{align}
& \sum_{a}\frac{\rho _{a}}{2}(\chi _{a}^{\prime }-\mathcal{A})^{2}=\frac{%
\rho }{2}(\chi ^{\prime }-\mathcal{A})^{2}+\frac{K_{\phi }}{2}\left( \phi
^{\prime }\right) ^{2}, \\
\qquad & K_{\phi }=\frac{\rho _{1}\rho _{2}}{\rho _{1}+\rho _{2}},  \notag
\label{eq:kin_diag_SM}
\end{align}%
and the Lifshitz term separates in the same basis,
\begin{equation}
\sum_{a}\eta _{a}(\chi _{a}^{\prime }-\mathcal{A})=\eta _{c}(\chi ^{\prime }-%
\mathcal{A})+\eta _{\phi }\phi ^{\prime },  \label{eq:lifshitz_split_SM}
\end{equation}%
with
\begin{equation}
\eta _{c}=\eta _{1}+\eta _{2},\quad \eta _{\phi }=\eta _{1}c_{2}-\eta
_{2}c_{1}=\frac{\eta _{1}\rho _{2}-\eta _{2}\rho _{1}}{\rho _{1}+\rho _{2}}.
\label{eq:eta_split_SM}
\end{equation}%
The relative-phase part therefore vanishes if the band-normalized
magneto-electric responses are equal, $\eta _{1}/\rho _{1}=\eta _{2}/\rho_{2}$. A nonzero topological bias requires the band-selective condition, $\eta _{1}/\rho _{1}\neq \eta _{2}/\rho _{2}$.

The London functional becomes
\begin{align}
\mathcal{G}=\int_{0}^{2\pi }\!\mathrm{d}\varphi \Bigg[& \frac{\rho }{2}(\chi
^{\prime }-\mathcal{A})^{2}+\eta _{c}(\chi ^{\prime }-\mathcal{A})  \notag \\
& +\frac{K_{\phi }}{2}(\phi ^{\prime 2}+\eta _{\phi }\phi ^{\prime
}-g_{1}\cos \phi -g_{2}\cos 2\phi \Bigg]  \notag \\
& +\frac{1}{8\pi }\!\int \!\mathrm{d}^{2}r\,(\bm h-\bm H)^{2}.
\label{eq:GL_sep_SM}
\end{align}%
Thus, the local energy density separates into common and relative parts. The corresponding sectors remain coupled only through the global winding constraints and the subsequent minimization with respect to the fluxoid sectors.

Turning to the common-phase sector, completing the square yields
\begin{equation}
\frac{\rho }{2}(\chi ^{\prime }-\mathcal{A})^{2}+\eta _{c}(\chi ^{\prime }-%
\mathcal{A})=\frac{\rho }{2}\left( \chi ^{\prime }-\mathcal{A}+\frac{\eta
_{c}}{\rho }\right) ^{2}-\frac{\eta _{c}^{2}}{2\rho }.
\label{eq:common_complete_SM}
\end{equation}%
The last term gives an $n$-independent constant after integration over the ring, therefore this constant term is dropped. Therefore, the common-phase Lifshitz term enters only as a rigid flux offset,
\begin{equation}
f_{L}\equiv \frac{\eta _{c}}{\rho }.  \label{eq:fL_def_SM}
\end{equation}%
The common phase couples to the gauge field as in the single-band model, but the integer windings pertain to the individual condensate phases rather than to the stiffness-weighted combination $\chi $. Single-valuedness of each
order-parameter component imposes conditions
\begin{equation}
\frac{1}{2\pi }\oint \chi _{1}^{\prime }\,\mathrm{d}\varphi =n_{1},\quad
\frac{1}{2\pi }\oint \chi _{2}^{\prime }\,\mathrm{d}\varphi =n_{2},\quad
n_{1},n_{2}\in \mathbb{Z},  \label{eq:fluxoid_SM}
\end{equation}%
so that the relative phase winds with integer $n=n_{1}-n_{2}$, while the common phase winds so that
\begin{equation}
\nu _{\chi }=\frac{1}{2\pi }\oint \chi ^{\prime }\,\mathrm{d}\varphi
=c_{1}n_{1}+c_{2}n_{2}=n_{1}-c_{2}\,n,  \label{eq:nu_chi_SM}
\end{equation}%
which is generally a 
non-integer. At fixed integer relative winding $n$, minimizing the shifted common stiffness $\tfrac{\rho }{2}(\chi ^{\prime }-\mathcal{A}+f_{L})^{2}$, together with the screening (magnetic-field) energy, with respect to the gauge field and the
common fluxoid sector is therefore equivalent to minimizing over the integer $n_{1}$. For the thin-walled ring of height $L$, mean radius $R$, and wall thickness $d\ll \lambda $, this is the self-consistent calculation performed in Ref. \cite{Kuplevakhsky2011_SM}. In the weak-screening regime, with $\epsilon \equiv dR/2\lambda ^{2}\ll 1$, it yields a parabolic energy with respect to the gauge-invariant flux mismatch, with the curvature set by the electromagnetic scale $E_{\ast }=\Phi _{0}^{2}\epsilon /2L_{m}$ and geometric inductance $L_{m}=4\pi ^{2}R^{2}/L$. To carry out the minimization explicitly, we adopt
the doubly-connected gauge convention $\tfrac{1}{2\pi }\oint \mathcal{A}\,\mathrm{d}\varphi =-f_{H}$, for which the gauge-invariant mismatch vanishes at integer values of the reduced flux, after the minimization with respect to the common-phase fluxoid integer. For the uniform magneto-electric coefficient assumed here, $f_{L}$ is independent of the angular coordinate $\varphi $, and the gauge-invariant winding mismatch of the shifted common-phase sector~\eqref{eq:common_complete_SM} is
\begin{equation}
\begin{split}
M(n_{1};n)& =\frac{1}{2\pi }\oint \!\big(\chi ^{\prime }-\mathcal{A}+f_{L}%
\big)\mathrm{d}\varphi \\
& =n_{1}-c_{2}n+f_{H}+f_{L},
\end{split}
\label{eq:mismatch_SM}
\end{equation}%
where $\nu _{\chi }=n_{1}-c_{2}n$ is taken from Eq.~\eqref{eq:nu_chi_SM}.
The self-consistent London minimization of Ref.~\cite{Kuplevakhsky2011_SM} renders the field energy quadratic with respect to the mismatch, $G_{\mathrm{%
em}}^{L}=E_{\ast }(1-\epsilon )\,M^{2}$. At fixed relative winding $n$ the only remaining free integer is the common winding $n_{1}$. Since $M$ is linear with respect to $n_{1}$, with unit coefficient,
\begin{equation}
\begin{split}
\min_{n_{1}\in \mathbb{Z}}M(n_{1};n)^{2}& =\big[(f_{H}-nc_{2}+f_{L})-\lfloor
f_{H}-nc_{2}+f_{L}\rceil \big]^{2} \\
& =h^{2}(f_{H}-nc_{2}+f_{L}),
\end{split}
\label{eq:min_n1_SM}
\end{equation}%
the optimum being $n_{1}=-\lfloor f_{H}-nc_{2}+f_{L}\rceil $. Minimizing over $n_{1}$ thus folds the parabola into its periodic envelope, the sawtooth, written via $h(x)\equiv x-\lfloor x\rceil$, the signed deviation of $x$ from the nearest integer, and term$-c_{2}n$, inherited from the common-phase winding~\eqref{eq:nu_chi_SM}, enters its argument, giving the
electromagnetic contribution in the relative-phase sector:
\begin{equation}
G_{\mathrm{em}}^{L}(n;f_{H})=E_{\ast }(1-\epsilon )\,h^{2}\!\left(
f_{H}-nc_{2}+f_{L}\right) ,  \label{eq:Gem_SM}
\end{equation}%
with $f_{H}=\Phi _{H}/\Phi _{0}$ being the reduced applied flux. The sign in front of $c_{2}n$ reflects two conventions: the definition $n=n_{1}-n_{2}$ and the flux orientation in Eq.~\eqref{eq:mismatch_SM}. Flipping either convention maps $n\to -n$, i.e., it interchanges the labels of the two chiralities, while the set of branch energies remains unchanged.

In the relative-phase sector, if the respective magneto-electric coefficient is azimuthally uniform, as assumed in the radial-Rashba geometry adopted in the main text, the relative Lifshitz term is purely topological. Since the relative phase winds by $2\pi n$,
\begin{equation}
\int_{0}^{2\pi }\!\mathrm{d}\varphi \;\eta _{\phi }\phi ^{\prime }=2\pi \eta
_{\phi }n\equiv E_{\ast }\,\Lambda n,\qquad \Lambda \equiv \frac{2\pi \eta
_{\phi }}{E_{\ast }}.  \label{eq:Lambda_SM}
\end{equation}%
The remaining energy in the relative-phase sector is
\begin{align}
E_{\mathrm{rel}}[\phi ]& =\int_{0}^{2\pi }\mathrm{d}\varphi \left[ \frac{%
K_{\phi }}{2}(\phi ^{\prime 2}+V(\phi )\right] , \\
\qquad V(\phi )& =-g_{1}\cos \phi -g_{2}\cos 2\phi .  \notag
\end{align}%
Let $E_{0}(|n|)$ denote the lowest stationary solution with winding number $|n|$ of the corresponding double-sine-Gordon equation. The dimensionless soliton self-energy used in the main text is
\begin{align}
Y_{\mathrm{sol}}^{\mathrm{DSG}}(|n|)& =\frac{E_{0}(|n|)-2\pi V_{\min }}{%
E_{\ast }}, \\
\qquad V_{\min }& =\min_{\phi }V(\phi ),\qquad Y_{\mathrm{sol}}^{\mathrm{DSG}%
}(0)=0.  \notag
\end{align}%
Collecting the electromagnetic energy, the soliton self-energy, and the topological Lifshitz bias gives the reduced Gibbs energy
\begin{equation}
g(n;f_{H})=(1-\epsilon )h^{2}(f_{H}-nc_{2}+f_{L})+Y_{\mathrm{sol}}^{\mathrm{%
DSG}}(|n|)+\Lambda n.  \label{eq:gtotal_SM}
\end{equation}

\section*{II. The running soliton: the first integral and the elliptic solution}

The relative-phase energy on the ring is
\begin{align}
& E[\phi ]=\int_{0}^{2\pi }\!\mathrm{d}\varphi \left[ \tfrac{1}{2}K_{\phi
}(\phi ^{\prime 2}+V(\phi )\right] ,  \notag \\
& V(\phi )=-g_{1}\cos \phi -g_{2}\cos 2\phi ,  \label{eq:Erel_SM}
\end{align}%
whose Euler--Lagrange equation is the double sine-Gordon equation, $-K_{\phi }\phi ^{\prime \prime }+g_{1}\sin \phi +2g_{2}\sin 2\phi =0$, with the first integral
\begin{equation}
\tfrac{1}{2}K_{\phi }(\phi ^{\prime 2}=\mathcal{E}+V(\phi )=\mathcal{E}%
-g_{1}\cos \phi -g_{2}\cos 2\phi .  \label{eq:firstint_SM}
\end{equation}%
A running solution carries winding\ number $|n|$, i.e.\ $\phi $ increases by $2\pi |n|$ over one circuit. This requires the right-hand side of Eq.~\eqref{eq:firstint_SM} to remain strictly positive, i.e.,
\begin{equation}
\mathcal{E}>-V_{\min },\qquad V_{\min }=\min_{\phi }V(\phi ).
\label{eq:cond_SM}
\end{equation}%
With the substitution $y=\tan ^{2}(\phi /2)$, Eq.~\eqref{eq:firstint_SM} maps to a quartic. Using $\cos \phi =(1-y)/(1+y)$ and $\cos 2\phi=(1-6y+y^{2})/(1+y)^{2}$, the turning-point polynomial is
\begin{equation}
P(y)=(\mathcal{E}+g_{1}-g_{2})\,y^{2}+(2\mathcal{E}+6g_{2})\,y+(\mathcal{E}%
-g_{1}-g_{2}).  
\label{eq:quartic_SM}
\end{equation}

\subsection{A. The real-root regime ($g_{2}\geq 0$)}

When the discriminant of $P$ is positive---which includes the entire $g_{2}\geq 0$ half-plane---the two roots are real and negative, $P(y)=(\mathcal{E}+g_{1}-g_{2})(y+\alpha ^{2})(y+\beta ^{2})$ with $\beta >\alpha>0 $. Here $y=-\alpha ^{2}$ and $y=-\beta ^{2}$ are two roots of the turning-point polynomial~\eqref{eq:quartic_SM}, so that $\alpha $ and $\beta
$ are fixed by $\mathcal{E}$, $g_{1}$, and $g_{2}$ through the Vieta relations
\begin{equation}
\alpha ^{2}+\beta ^{2}=\frac{2\mathcal{E}+6g_{2}}{\mathcal{E}+g_{1}-g_{2}}%
,\qquad \alpha ^{2}\beta ^{2}=\frac{\mathcal{E}-g_{1}-g_{2}}{\mathcal{E}%
+g_{1}-g_{2}}.  \label{eq:vieta_SM}
\end{equation}%
Because both roots are negative, they correspond to no real value of $y=\tan^{2}(\phi /2)$, consistent with the monotonically advancing (running) phase; $\alpha $ sets the amplitude scale of the solution below and, together with $\beta $, fixes the elliptic modulus $m=1-\alpha ^{2}/\beta ^{2}$ that controls the soliton width. The running solution is
\begin{equation}
\tan \frac{\phi (\varphi )}{2}=\alpha \,\mathrm{sc}\!\big[\Omega (\varphi
-\varphi _{0})\,\big|\,m\big],\qquad m=1-\frac{\alpha ^{2}}{\beta ^{2}},
\label{eq:sol_SM}
\end{equation}%
where $\mathrm{sc}=\mathrm{sn}/\mathrm{cn}$ and $\varphi _{0}$ is the soliton's centre. The unwrapped phase is obtained by continuing through the poles of $\mathrm{sc}$. Each period $2K(m)$ of the argument adds $2\pi $ to $\phi $, where $K(m)$ is the complete elliptic integral of the first kind.
The wavenumber and the ring quantization condition are
\begin{equation}
\Omega =\beta \sqrt{\frac{\mathcal{E}+g_{1}-g_{2}}{2K_{\phi }}},\qquad 2\pi
\Omega =2|n|K(m),  \label{eq:quant_SM}
\end{equation}%
which fixes $\mathcal{E}$ (hence $\alpha ,\beta ,m,\Omega $) uniquely for
each $|n|$.

\subsection{B. The complex-root regime ($g_{2}<0$)}

For $g_2<0$ the discriminant of $P$ can become negative in a region confined to the lower half-plane; the roots of $P$ are then complex conjugate, $P(y)=(\mathcal{E}+g_1-g_2)\,[(y+p)^2+q^2]$. The running solution is the corresponding $\mathrm{cn}$-type elliptic function,
\begin{equation}
\tan\frac{\phi(\varphi)}{2}=\frac{a+b\,\mathrm{cn}(\Omega^{\prime
}(\varphi-\varphi_0)\,|\,\tilde m)} {1+\mathrm{cn}(\Omega^{\prime
}(\varphi-\varphi_0)\,|\,\tilde m)},  \label{eq:cnsol_SM}
\end{equation}
with constants $a,b,\Omega^{\prime },\tilde m$ determined by $p,q$ through the standard quadratic-to-elliptic reduction, whose explicit form is not needed here. The same quantization $2\pi\Omega^{\prime }=2|n|K(\tilde m)$ applies. In practice we evaluate $E_0(|n|)$ in this regime directly by the quadrature from Eq.~\eqref{eq:firstint_SM}, which is solver-independent and agrees with the boundary-value solution of the DSG equation to better than $10^{-7}$.

\emph{The branch selection.} In the double-well region ($g_{2}<0$, $|g_{1}|<-4g_{2}$) more than one stationary winding configuration of a given $|n|$ may coexist---a more uniformly wound branch and more localized kink-like branches. We obtain all converged running branches---analytically where the quartic \eqref{eq:quartic_SM} has real roots, and by a multi-start
boundary-value solver of the double sine-Gordon equation elsewhere---and take $E_{0}(|n|)$ to be the lowest, i.e.\ the thermodynamically selected one. Each running solution retains the translational zero mode $\varphi _{0}$ and the twofold chirality $\phi \rightarrow -\phi $. It is the latter degeneracy, present at $\Lambda =0$, that the relative Lifshitz term lifts.

\section*{III. Harmonic integrals $I_1$ and $I_2$}

The soliton energy quoted in the main text,
\begin{equation}
E_{0}(|n|)=2\pi \mathcal{E}-2g_{1}I_{1}-2g_{2}I_{2},  \label{eq:E0_SM}
\end{equation}%
requires
\begin{equation}
I_{1}=\int_{0}^{2\pi }\!\cos \phi \,\mathrm{d}\varphi ,\qquad
I_{2}=\int_{0}^{2\pi }\!\cos 2\phi \,\mathrm{d}\varphi .
\label{eq:I12_def_SM}
\end{equation}%
In the real-root regime we use $\cos \phi =(1-y)/(1+y)$ with $y=\alpha ^{2}%
\mathrm{sc}^{2}=\alpha ^{2}\mathrm{sn}^{2}/\mathrm{cn}^{2}$. Writing $%
s\equiv \mathrm{sn}^{2}(\Omega (\varphi -\varphi _{0})|m)$,
\begin{equation}
\cos \phi =\frac{1-(1+\alpha ^{2})s}{1-(1-\alpha ^{2})s}=C+\frac{D}{1-q\,s},
\label{eq:cosphi_SM}
\end{equation}%
with the constants
\begin{equation}
C=\frac{1+\alpha ^{2}}{1-\alpha ^{2}},\quad D=-\frac{2\alpha ^{2}}{1-\alpha
^{2}},\quad q=1-\alpha ^{2}.  \label{eq:CDq_SM}
\end{equation}%
Changing variables to $u=\Omega (\varphi -\varphi _{0})$ (so that $\mathrm{d}\varphi =\mathrm{d}u/\Omega $ and the circuit $\varphi :0\rightarrow 2\pi $ maps to $u:0\rightarrow 2|n|K(m)$) and using the standard result $\int_{0}^{2K}\!\mathrm{d}u/(1-q\,\mathrm{sn}^{2}u)=2\,\Pi (q;m)$, where $\Pi$ is the complete elliptic integral of the third kind, one obtains the
closed form
\begin{equation}
I_{1}=\frac{2|n|}{\Omega }\Big[\,C\,K(m)+D\,\Pi (q;m)\,\Big]\,,
\label{eq:I1_SM}
\end{equation}%
which we have verified against direct numerical integration. For $I_{2}$ we use $\cos (2\phi )=2\cos ^{2}\phi -1$ with Eq.~\eqref{eq:cosphi_SM},
\begin{align}
& \cos 2\phi =2\Big(C+\frac{D}{1-q\,s}\Big)^{2}-1  \notag \\
& =(2C^{2}-1)+\frac{4CD}{1-q\,s}+\frac{2D^{2}}{(1-q\,s)^{2}},
\label{eq:cos2phi_SM}
\end{align}%
so that, with the additional standard integral $\int_{0}^{2K}\!\mathrm{d}%
u/(1-q\,\mathrm{sn}^{2}u)^{2}=2\,\mathcal{P}_{2}(q;m)$, where
\begin{align}
\mathcal{P}_{2}(q;m)=& \frac{1}{2(q-1)(m-q)}\Big[\,(m-q)K(m)+q\,E(m)  \notag
\\
& +(2q-3m-q^{2}+2qm)\,\Pi (q;m)\Big],  \label{eq:P2_SM}
\end{align}%
one finds
\begin{equation}
\begin{split}
I_{2}=\frac{2|n|}{\Omega }\Big[& (2C^{2}-1)K(m)+4CD\,\Pi (q;m) \\
& +2D^{2}\,\mathcal{P}_{2}(q;m)\Big].
\end{split}
\label{eq:I2_SM}
\end{equation}%
Eqs.~\eqref{eq:I1_SM} and~\eqref{eq:I2_SM}, inserted into Eq.~\eqref{eq:E0_SM}, give the soliton self-energy $Y_{\mathrm{sol}}^{\mathrm{DSG}}(|n|)=[E_{0}(|n|)-2\pi V_{\min }]/E_{\ast }$, evaluated at the benchmark stiffness $\kappa _{\phi }\equiv \pi K_{\phi }/E_{\ast }=c_{1}c_{2}$, used
to compute all figures of the main text. In the complex-root regime the analogous expressions follow from Eq.~\eqref{eq:cnsol_SM}. In practice $E_{0}(|n|)$ is obtained there by quadrature, as noted above.

\section*{IV. The localized kink versus delocalized winding solutions}

Within the finite-winding (soliton) region the texture interpolates between two limits, distinguished by the kink half-width
\begin{equation}
\ell_\varphi=\sqrt{\frac{K_\phi}{\Delta V}}, \qquad w=R\,\ell_\varphi, \qquad \Delta V=V_{\max}-V_{\min},
\label{eq:ell_SM}
\end{equation}
compared with the full angular circuit $2\pi $; the corresponding physical width is $w=R\ell_\varphi $, to be compared with the circumference $2\pi R$.

\emph{Strong locking} ($\ell_\varphi \ll 2\pi $, away from the origin of the $(g_{1},g_{2})$ plane). The potential is deep, with the modulus $m\rightarrow 1$, and the elliptic solution~\eqref{eq:sol_SM} reduces to a well-separated
train of localized $2\pi $ kinks: $\phi $ sits near a minimum of $V$ over most of the ring and advances through narrow domain walls of width $\sim \ell_\varphi $. This is the genuine double--sine-Gordon soliton.

\emph{Weak locking} ($\ell_\varphi\gg 2\pi$, near $g_1=g_2=0$). The potential is too shallow to localize a kink on the scale of the ring. The modulus $m\to0$, $\mathrm{sc}\to\tan$, and the solution~\eqref{eq:sol_SM} reduces to uniform winding, $\phi(\varphi)\simeq n\varphi$; a finite relative-phase
momentum state rather than a domain wall. In this limit $E_0(|n|)-2\pi V_{\min}\to\pi K_\phi n^2$ and the soliton self-energy reduces to the pure-winding kinetic cost $Y_{\mathrm{sol}}^{\mathrm{DSG}}(n)\to c_1c_2 n^2$.

The transition between these textures is a smooth crossover controlled by $m$ (equivalently by $\ell_\varphi/2\pi R$), not a thermodynamic phase boundary: $E_0(|n|)$ and all thermodynamic quantities are analytic across it. For this reason it is not drawn as a separate phase in the main-text phase diagram,
where the soliton region near the origin is of the delocalized,
finite-momentum type and the strongly-locked periphery is of the
localized-kink type.

\section*{V. The small-amplitude limit: the Leggett mode}

The same relative-phase sector that supports the static $2\pi$ soliton also carries the Leggett collective mode~\cite{Leggett1966_SM}, its small-amplitude counterpart. Writing $\phi=\phi_{\min}+u(\varphi,t)$ about a
homogeneous minimum $\phi_{\min}$ of $V$ and retaining the relative-phase inertia $\chi_\phi$, the dynamics of the
relative phase linearizes to
\begin{equation}
\chi_\phi\,\ddot u-K_\phi\,u^{\prime \prime }+V^{\prime \prime
}(\phi_{\min})\,u=0,  \label{eq:leggett_lin_SM}
\end{equation}
with $V^{\prime \prime }(\phi)=g_1\cos\phi+4g_2\cos2\phi$. The uniform ($q=0$%
) oscillation is the gapped Leggett mode,
\begin{equation}
\omega_L^2=\frac{V^{\prime \prime }(\phi_{\min})}{\chi_\phi},
\label{eq:leggett_gap_SM}
\end{equation}
fixed by the curvature of the locking potential at its minimum. In the single-harmonic case ($g_2=0$, $\phi_{\min}=0$) this reduces to $\omega_L^2=g_1/\chi_\phi\propto\Gamma_1$, the familiar result that the Leggett gap scales with the interband Josephson coupling. The Leggett mode and the interband phase soliton originate from the same relative-phase degree of freedom, but belong to different topological sectors: the Leggett mode is a small-amplitude dynamical excitation about a homogeneous $n=0$ minimum, whereas the soliton is a static finite-winding configuration with $n\neq0$. Within the finite-winding family, the elliptic modulus $m$ controls the crossover from the delocalized, nearly uniform winding ($m\to0$) to the localized-kink limit ($m\to1$); this crossover should not be identified with the small-amplitude Leggett limit. Consistently, the Leggett gap closes, $\omega_L\to0$, exactly at $g_1=g_2=0$, where the relative phase becomes a free (Goldstone) mode.

\section*{VI. The single-harmonic limit}

For $g_{2}=0$ the potential reduces to $V(\phi )=-g_{1}\cos \phi $ and the DSG equation becomes the sine-Gordon equation. The polynomial~\eqref{eq:quartic_SM} has $\mathcal{E}+g_{1}>0$ and $\mathcal{E}-g_{1}$ as the relevant combinations, and the running solution~\eqref{eq:sol_SM} becomes the standard sine-Gordon running solution with
\begin{equation}
\Omega =\sqrt{\frac{\mathcal{E}+g_{1}}{2K_{\phi }}}\,\frac{1}{\sqrt{m}}%
,\qquad 2\pi \Omega =2|n|K(m),  \label{eq:SG_quant_SM}
\end{equation}%
and modulus fixed by $m=2g_{1}/(\mathcal{E}+g_{1})$. The harmonic integral $I_{2}$ no longer enters $E_{0}$, and Eq.~\eqref{eq:E0_SM} reduces to $E_{0}(|n|)=2\pi \mathcal{E}-2g_{1}I_{1}$, recovering the exact sine-Gordon
soliton energy of Ref.~\cite{Kuplevakhsky2011_SM}. This provides an independent analytic check of the general double sine-Gordon expressions above in the $g_{2}\rightarrow 0$ limit.

\section*{VII. Order-of-magnitude estimates for the magneto-electric soliton bias}

The stabilization of a finite-winding soliton is controlled by the scaled odd-in-winding Lifshitz bias
\begin{equation}
\Lambda =\frac{2\pi \eta _{\phi }}{E_{\ast }}.
\end{equation}%
It is useful to express this in terms of the preferred relative-phase winding
\begin{equation}
q_{L}\equiv \frac{\eta _{\phi }}{K_{\phi }},
\end{equation}%
which is dimensionless because the phase coordinate on the ring is the azimuthal angle. Using
\begin{equation}
\kappa _{\phi }\equiv \frac{\pi K_{\phi }}{E_{\ast }},
\end{equation}%
one obtains the general relation
\begin{equation}
\Lambda =2\kappa _{\phi }q_{L}.  \label{eq:Lambda_qL_SM}
\end{equation}%
In the normalization used for the figures in the main text, $\kappa _{\phi}=c_{1}c_{2}$, and for the representative symmetric case $c_{1}=c_{2}=1/2$ this gives $\Lambda =q_{L}/2$. Thus the value $\Lambda \simeq 0.3$, for which a sizable soliton ground-state region appears in Fig.~2 of the main text, corresponds to
\begin{equation}
q_{L}\simeq 0.6,
\end{equation}%
i.e.\ the magneto-electric term need only favor a fraction of a $2\pi $ relative-phase twist around the ring.

The corresponding physical helical wave vector is $Q_{L}=q_{L}/R$, with $R$
the ring radius, so that
\begin{equation}
Q_{L}^{\mathrm{req}}\simeq 6\times 10^{5}\,\mathrm{m}^{-1}\left( \frac{q_{L}%
}{0.6}\right) \left( \frac{1\,\mathrm{\mu m}}{R}\right) ,
\label{eq:QL_req_SM}
\end{equation}%
and the associated helical length is
\begin{equation}
\ell_{L}=\frac{2\pi }{Q_{L}}\simeq 10\,\mathrm{\mu m}\left( \frac{0.6}{q_{L}%
}\right) \left( \frac{R}{1\,\mathrm{\mu m}}\right) .
\end{equation}%
This scale is mesoscopic rather than microscopic, which is the basic reason the stabilization condition is experimentally plausible.

A simple estimate of the field-induced magneto-electric wave vector may be written as
\begin{equation}
Q_L^{\mathrm{ME}} \sim \gamma_{\mathrm{rel}}\, \frac{\mu_B B_{\mathrm{ME}}}{%
\hbar v_F},  \label{eq:QME_est_SM}
\end{equation}
where $B_{\mathrm{ME}}$ is the field component that activates the
magneto-electric Lifshitz invariant, $v_F$ a representative Fermi velocity, and the dimensionless coefficient $\gamma_{\mathrm{rel}}$ is an effective phenomenological parameter summarizing the antisymmetric spin-orbit
coupling, the band selectivity of the magneto-electric response, and possible geometric reduction factors. Heuristically, $\gamma_{\mathrm{rel}}$ may be expected to track the spin-orbit splitting relative to the Fermi energy, weighted by a band-asymmetry factor; in particular the relative bias vanishes when the two band-normalized responses coincide,
\begin{equation}
\frac{\eta_1}{\rho_1}=\frac{\eta_2}{\rho_2},
\end{equation}
so the relevant microscopic quantity is not the total magneto-electric response but the difference between the two band-normalized responses.

Combining Eqs.~\eqref{eq:QL_req_SM} and~\eqref{eq:QME_est_SM}, and for absorbing factors of the order of unity into $\gamma _{\mathrm{rel}}$, the field required to reach a given $q_{L}$ is estimated as
\begin{equation}
B_{\mathrm{ME}}^{\mathrm{req}}\sim \frac{q_{L}}{0.879\,\gamma _{\mathrm{rel}}%
}\left( \frac{v_{F}}{10^{5}\,\mathrm{m/s}}\right) \left( \frac{1\,\mathrm{%
\mu m}}{R}\right) \,\mathrm{T},  \label{eq:Breq_SM}
\end{equation}%
so that for $q_{L}=0.6$,
\begin{equation}
B_{\mathrm{ME}}^{\mathrm{req}}\sim 6.8\,\mathrm{T}\left( \frac{0.1}{\gamma _{%
\mathrm{rel}}}\right) \left( \frac{v_{F}}{10^{5}\,\mathrm{m/s}}\right)
\left( \frac{1\,\mathrm{\mu m}}{R}\right) .
\end{equation}%
A conventional metal with $v_{F}\sim 10^{5}\,\mathrm{m/s}$ and a modest relative coefficient $\gamma _{\mathrm{rel}}\sim 0.1$ would thus require fields of several Tesla, whereas heavy-fermion or oxide-interface systems with smaller $v_{F}$, larger spin-orbit band selectivity, or larger ring radius reach the same regime at substantially lower fields.

Representative estimates are collected in Table~\ref{tab:platform_estimates}. These should be read as order-of-magnitude indicators rather than material-specific predictions, since $\gamma_{\mathrm{rel}}$ depends on the
microscopic band structure, orbital content, disorder, and device geometry. In practice, $B_{\mathrm{ME}}$ should be viewed as a slowly varied bias field that sets $\Lambda$, while the reduced flux $f_H$ can be tuned modulo one flux quantum by a local coil or flux line; the theory depends only on the resulting values of $f_H$ and $\Lambda$.

\begin{table*}[t]
\caption{Order-of-magnitude estimates of the magneto-electric field needed to reach $q_L\simeq0.6$, corresponding to $\Lambda\simeq0.3$ for $c_1=c_2=1/2 $ and $\protect\kappa_\protect\phi=c_1c_2$. The estimates use
Eq.~ \eqref{eq:Breq_SM} with $R=1\,\protect\mu\mathrm{m}$. The quoted $v_F$ and $\protect\gamma_{\mathrm{rel}}$ intervals are representative assumed ranges for each platform class, not material-specific fits.}
\label{tab:platform_estimates}%
\begin{ruledtabular}
\begin{tabular}{lccc}
Platform & $v_F$ (m/s) & $\gamma_{\rm rel}$ & $B_{\rm ME}^{\rm req}$ (T) \\
\hline
Heavy-fermion NCS, e.g.\ CePt$_3$Si
& $10^4$--$5\times10^4$ & $0.1$--$0.5$ & $0.1$--$3$ \\
Strong-SOC bulk NCS, e.g.\ Li$_2$Pt$_3$B
& $5\times10^4$--$2\times10^5$ & $0.2$--$0.5$ & $0.7$--$7$ \\
Moderate-SOC bulk NCS, e.g.\ Mo$_3$Al$_2$C
& $\sim10^5$ & $0.02$--$0.1$ & $7$--$34$ \\
Oxide interface, e.g.\ LaAlO$_3$/SrTiO$_3$
& $10^4$--$5\times10^4$ & $0.1$--$1$ & $0.1$--$3$
\end{tabular}
\end{ruledtabular}
\end{table*}

Bulk noncentrosymmetric multiband superconductors such as CePt$_3$Si, Li$_2$Pt$_3$B, and Mo$_3$Al$_2$C~\cite{Bauer2004_SM,Yuan2006_SM,BauerMo3Al2C2010_SM,Smidman2017_SM,BauerSigrist2012_SM} establish that antisymmetric spin-orbit coupling and band-dependent Fermi-surface properties are realistic ingredients, and are therefore natural candidates for the unequal band-normalized responses $\eta_1/\rho_1\ne\eta_2/\rho_2$ required for a nonzero relative Lifshitz bias. For the specific ring geometry of the main text, oxide-interface superconductors are particularly attractive: in LaAlO$_3$/SrTiO$_3$-based two-dimensional superconductors the Rashba coupling is interface-induced, gate tunable, and strongly affected by the multi-orbital $t_{2g}$ band structure~\cite{Reyren2007_SM,Caviglia2010_SM}. Patterned annular
oxide-interface devices, possibly supplemented by side gates or
side-wall/interface engineering, could provide a route toward the local radial inversion-breaking geometry assumed in the main text. Even when the interface normal is fixed, electrostatic gating offers a practical way to tune the orbital occupation and the band selectivity of the magneto-electric response.

In the crystal frame the invariant reads $\sum_{a}K_{ij}^{(a)}B_{i}j_{a,j}$, with $K_{ij}^{(a)}$ the magneto-electric (gyrotropic) tensor fixed by the noncentrosymmetric point group; $\kappa _{a}$ is the projection $K_{\varphi z}^{(a)}$ relevant here. A uniform term around the ring requires this expression to be azimuth-independent, which holds when the local inversion-breaking axis follows the radial direction (e.g.\ an interface-induced Rashba field normal to the circumference), for which $(\hat{\mathbf{r}}\times \hat{\mathbf{z}})\cdot \hat{\boldsymbol{\varphi }}=-1$
around the entire ring.

Finally, the bias fields of Table~\ref{tab:platform_estimates} must be compatible with superconductivity in the ring. In the cylindrical geometry, an axial bias field lies parallel to the thin wall, whose normal is radial, so orbital depairing across the wall thickness is governed by the thin-film
parallel-field scale
\begin{equation}
H_{c\parallel}\simeq \frac{\sqrt{12}\,\Phi_0}{2\pi\,\xi d}.
\label{eq:Hcpar_SM}
\end{equation}
For $d\lesssim\xi\sim10$--$20\,\mathrm{nm}$, this gives a scale of a few to ten tesla, increasing as the wall is thinned~\cite{Tinkham_SM}. Paramagnetic depairing provides an additional constraint, but in noncentrosymmetric superconductors antisymmetric spin-orbit coupling can strongly modify the
Pauli limit, and in several materials, including CePt$_3$Si, the measured upper critical fields exceed the weak-coupling Pauli estimate \cite{Bauer2004_SM,Smidman2017_SM}. Thus the same spin-orbit physics that allows the magneto-electric response can also enlarge the field window in which the proposed bias may be applied, although the quantitative limit is material
and geometry dependent.

\end{document}